\documentclass[11pt, a4paper]{article}
\usepackage{graphicx}

\newcommand{\tri}{\triangle}

\newcommand{\sep}{ \ \ \ , \ \ \ }

\newcommand{\beq}{\begin{equation}}
\newcommand{\beqn}{\begin{eqnarray}}
\newcommand{\eeqn}{\end{eqnarray}}
\newcommand{\pp}{\partial}
\newcommand{\dd}{{\rm d}}

\newcommand{\eqs}{equations }
\newcommand{\fig}{figure }

\newcommand{\la}{\langle}
\newcommand{\ra}{\rangle}

\newcommand{\cF}{F}
\newcommand{\cI}{I}
\newcommand{\cJ}{J}
\newcommand{\cD}{{\cal D}}

\newcommand{\e}{{\mathbf E}}

\newcommand{\q}{L}
\newcommand{\M}{M}
\newcommand{\eeq}{\end{equation}}

\newcommand{\icn}{$i=1, \ldots, n$}

\newcommand{\mm}{{\cal{M}}}
\newcommand{\ito}{It\^{o}'s Lemma }
\newcommand{\eqq}{equation }

\begin{document}

\title{
Correlated multi-asset portfolio
optimisation 
\\
with transaction
cost}

\author{Siu Lung Law,$^{1}$\thanks{Present address: Citigroup Global Markets Asia Limited,
Citibank Plaza, 3 Garden Road, Central, Hong Kong} \ \ 
Chiu Fan Lee,$^{2}$\thanks{Author for correspondence (C.Lee1@physics.ox.ac.uk).}
\\
 Sam Howison$^{1}$ and Jeff N.\ Dewynne${^1}$ 
\\
\\
$^1${\it Oxford Centre for Industrial and Applied Mathematics}
\\
{\it Oxford University, Oxford OX1 3LB, UK}
\\
\\
$^2${\it Physics Department, Clarendon Laboratory}
\\
{\it Oxford University, Oxford OX1 3PU, UK}
}


\maketitle

\begin{abstract}
We employ perturbation analysis technique to study 
multi-asset portfolio optimisation with transaction cost. We allow for correlations in risky assets and obtain optimal trading methods for general utility functions. 
Our analytical results are supported by numerical simulations in the context of the Long Term Growth Model.
\end{abstract}


\maketitle


\section{Introduction}
\begin{figure}
\caption{A schematic diagram depicting the optimal trading strategy for the long term growth model. The optimal holding of the risky asset, $A^* (\propto \Pi)$, is shown 
by the red curve. The sale-no-transaction boundary is given by $(A^* + \alpha_+)$ and the purchase-no-transaction boundary is given by $(A^* + \alpha_-)$ where $\alpha_- 
<0$ (cf.  \eqq \ref{M2}). In this work, we assume that $\alpha_+ = -\alpha_-$.
} 
\label{pic}
\begin{center}
\includegraphics[scale=.55]{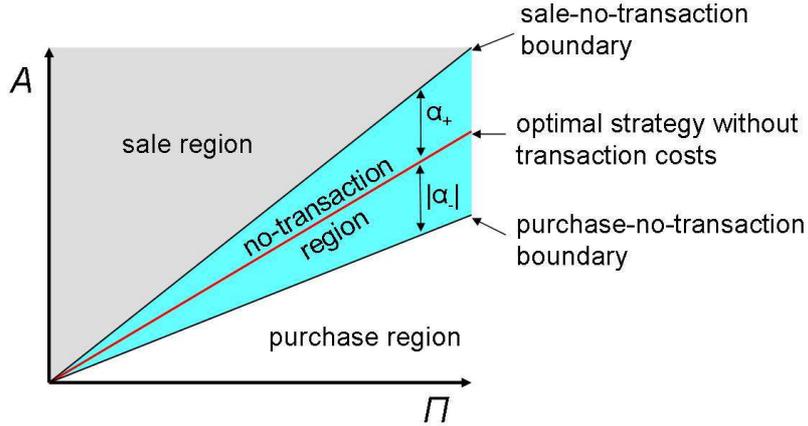}
\end{center}
\end{figure}

In recent years, the 
study of portfolio optimisation under non-zero transaction cost has 
received its due attention 
(Davis \& Norman 1990; Atkinson \& Wilmott 1995; Morton \& Pliska 1995; Akian {\it et al.} 1996; Atkins \& Dyl 1997; Atkinson \& Al-Ali 1997; Atkinson {\it et al.} 1997; 
Atkinson \& Mokkhavesa 2001, 2003, 2004; Mokkhavesa \& Atkinson 2002; Chellathurai \& Draviam 2005). In the literature, it is found that the incorporation of transaction 
fees into the model introduces a no-transaction region around the original optimal curve, surrounded by purchase and sale regions (cf.\ \fig \ref{pic}). 
Most of previous work focused on having only one risky asset (stock) and one
risk-free asset (bond), except the study by Atkinson and Mokkhavesa (2004) in which portfolio with multiple risky assets is analysed. In this work, 
the authors are able 
to obtain the optimal investment strategy with the assumption that the risky assets are uncorrelated. 
Here, we go beyond this restriction and consider correlated risky assets. By assuming that the purchase and sale boundaries are of an equal distance away from the 
optimal curve, we obtain analytical expressions for the optimal trading strategy for general utility functions. We further support our analytical results with numerical 
simulations in the context of the Long Term Growth Model.

The plan of the paper is as follows:
In \S2, we will consider portfolio optimisation without transaction costs, thereby introduce the dynamic programming method employed. We will then consider trading with 
transaction cost in \S3. The details of our simulation method in support of our analytical results are given in \ref{simulations}. For reference, the expressions for the 
derivatives of the value function in terms of the expansion parameter are given in \ref{appB}.

\section{Trading without transaction costs}

We consider a market with investment opportunities on
$n$ stocks and a risk free bond, and we let $A_i (t)$, $B(t)$ and $\Pi(t) \equiv B(t) +\sum_{i=1}^n A(t)$ be the values held
in stock $i$, the value held in risk free bond and the total wealth at time $t$ respectively.
 We
assume that $A_i(t)$ follows a geometric Brownian motion with growth rate
$\mu_i$ and volatility $\sigma_i$, and the risk free bonds, $B$,
compounds continuously with risk free rate $r$. The volatilities
$\sigma_i$, growth rates $\mu_i$ and interest rate $r$ are assumed to be
constant. Cash generated or needed from the purchase or sale of
stocks is immediately invested or withdrawn from the risk free
bonds. 
In the absence of transaction costs, the problem is easily solved without recourse to perturbation analysis and this section will serve to familiarise the readers with 
the use of dynamic programming method in this optimisation problem.

The market model equations are represented by the followings:
\begin{eqnarray}
\label{main}
\dd A_i & = & \mu _i A_i \dd t+ \sigma_i A_i \dd X_i\ ,\ \ i = 1, \cdots , n
\nonumber
\\
\dd B & = & rB \dd t 
 =  r\left(\Pi - \sum^n_{i=1} A_i \right) \dd t \nonumber
\\
\dd \Pi & = & rB\dd t+ \sum^n_{i=1} \mu_i A_i \dd t + \sum^n_{i=1} \sigma_i
A_i \dd X_i \ ,
\end{eqnarray}
where $X_i$ , \icn, are Weiner processes whose
correlations, $-1 \le \rho _{ij} \le 1$, are assumed constant.
At time $t=0$, an investor has an amount $\Pi(t=0)$ of
resources and the problem is to allocate investments over the 
time horizon $t \in [0, T]$, so as to maximise the following expectation value:
\[
{\mathbf E}  \left[ \cF(\Pi(T))+ \int_{0}^{T} \cI(\Pi(t')) dt' 
\right] \ .
\]
The functions $\cI$ and $\cF$ can represent anything from utility to the year end bonus of the trader. For example, if we assume that $\cI=0$ and 
$\cF(\Pi(T))=\log(\Pi(T))$, then the opimisation problem constitutes the Long Term Growth Model and the goal would then be to optimise the logarithm of the final wealth. 
To make financial sense, we will assume that the utility functions are increasing and concave down, i.e.,
\begin{eqnarray}
\frac{\pp \cI}{\pp \Pi} \geq  0 \ \ \ & ,& \ \ \ \frac{\pp ^2\cI}{\pp \Pi^2} \leq 0
\\
\frac{\pp \cF}{\pp \Pi}\geq  0 \ \ \ &,& \ \ \ \frac{\pp ^2\cF}{\pp \Pi^2} \leq 0 \ .
\end{eqnarray}

We restate the optimisation problem in dynamic programming form by first defining 
the
optimal expected value function, $\cJ(\Pi, t)$:
\begin{equation}
\label{no_transaction_costs}
\cJ(\Pi, t) \equiv  \max_{A_i} \e_t \left[  \cF(\Pi) +\int_{t}^{T} \cI(\Pi(t'))
d{t'} \right] \ .
\end{equation}
We now apply the Bellman Principle and \ito to the above value function to obtain the following Hamilton-Bellman-Jacobi equation (Kamien 1991):
\begin{eqnarray}
0&= & \max_{A_1, \ldots, A_n} \Bigg[ \cI +  \frac{\pp \cJ}{\pp t }+
r\left(\Pi- \sum^n_{i=1} A_i \right)
\frac{\pp \cJ}{\pp \Pi} \nonumber
\\
&&
+\sum^n_{i=1} \mu_i A_i \frac{\pp \cJ}{\pp \Pi}  +\frac{1}{2}
\sum^n_{i,j=1} \Omega_{ij} A_iA_j \frac{\pp^2 \cJ}{\pp \Pi^2} \Bigg] \ ,
\label{multibellman}
\end{eqnarray}
with the boundary condition
$\cJ(\Pi, T)  =  \cF(\Pi(T))$. In the above equation, $\Omega_{ij} \equiv \sigma_i \sigma_j \rho_{ij}$  is the 
standard covariance matrix, and $A_i$ act as the control parameters in the context of dynamic programming. 

In matrix notation, we
can rewrite \eqq (\ref{multibellman}) as
\begin{equation} 
\label{Sec1_HJB}
0 = \max_{ A}  \left[ \cJ_t + \cI + r \Pi \cJ_\Pi +
(\hat{\mu} \cdot A)\cJ_\Pi + \frac{1}{2} (A \cdot \Omega
A) \cJ_{\Pi \Pi} \right]\ .
\end{equation}
where symbols without an index  denote the corresponding vectors (e.g. $A = (A_1, \ldots, A_n)^T$). We have also introduced a new vector 
$\hat{\mu}$, which is defined to be $(\mu_1-r, \ldots, \mu_n-r)^T$.

By
differentiating \eqq (\ref{Sec1_HJB}) with respect to $A$, one obtains as the solution to the HBJ Equation:
\beq
\frac{\pp \cJ}{\pp \Pi}\hat{\mu}+ \frac{\pp^2 \cJ}{\pp \Pi^2}
\Omega A  =  0 \ .
\eeq
Therefore, the optimal portfolio corresponds to:
\begin{equation}
\label{A}
A^*  =  -  \frac{\pp \cJ}{\pp \Pi} \left( \frac{\pp^2 \cJ}{\pp \Pi^2} \right)^{-1}\Omega^{-1}
\hat{\mu}
\ . 
\end{equation}

\subsection{Example: the Long Term Growth Model}
In this model, our aim is to maximize $\e[\log \Pi (T)]$. The value function is thus
\beq
J(\Pi, t) = \max_{A} \e_t [\log (\Pi(T))] \ .
\eeq
such that $J(\Pi, T) = \log (\Pi(T))$. This boundary condition together with the differential equation obtained by substituting \eqq (\ref{A}) into \eqq (\ref{Sec1_HJB}) 
implies that 
\begin{equation}
\label{M1}
\cJ(\Pi,t) =  \log \Pi + \left(r + \frac{1}{ 2}  \hat{\mu} \cdot
\Omega^{-1} \hat{\mu}\right)(T-t)\ .
\end{equation}
The optimal portfolio from \eqq (\ref{A}) is therefore given by:
\beq
\label{M2}
A^* = \Pi \Omega^{-1}  \hat{\mu}\ .
\eeq
For the case of having two-risky assets, the optimal portfolio corresponds to having
$B^* = q\Pi $ with $q \equiv 1- {\rm Tr} [\Omega^{-1} \hat{\mu}]$, and $A_i^* = p_i \Pi$ with $p_i \equiv [\Omega^{-1} \hat{\mu}]_i$. By the It\^{o}'s lemma, we have
\beq
\dd (\log \Pi)=\left(rq+\mu_1 p_1 + \mu_2 p_2- \frac{\beta^2}{2} \right)\dd t +\beta \dd X
\eeq 
where
\beq
\beta = \sqrt{\sigma_1^2 p_1^2+2\sigma_1\sigma_2 p_1p_2 \rho_{12} +\sigma_2^2 p_2^2} \ .
\eeq
The optimal expected payoff in this model is therefore:
\beq 
\label{analytical}
\e[\log (\Pi(T))] = \e[\log (\Pi(0))] + 
\left(rq+\mu_1 p_1 + \mu_2 p_2- \frac{\beta^2}{2} \right)T \ .
\eeq

\begin{figure}
\caption{Analytical and simulation results for the two-risky-asset market with model parameters given in \ref{simulations}. {\it Left plot:} $Y_1 \equiv \la \log(\Pi(T)) 
\ra$ denotes the performance of the optimal trading strategy, while $Y_2$ denotes the performance of the sub-optimal strategy obtained with the correlation between the 
two risky assets ignored.
{\it Right plot:} The plots of $(Y_1 - Y_2)$ and the standard errors of the means for $Y_1$ and $Y_2$ versus time based on our simulations. Note that $S = 4000$ is the 
number of samples in the simulations.
} 
\label{sec1}
\begin{center}
\includegraphics[scale=.5]{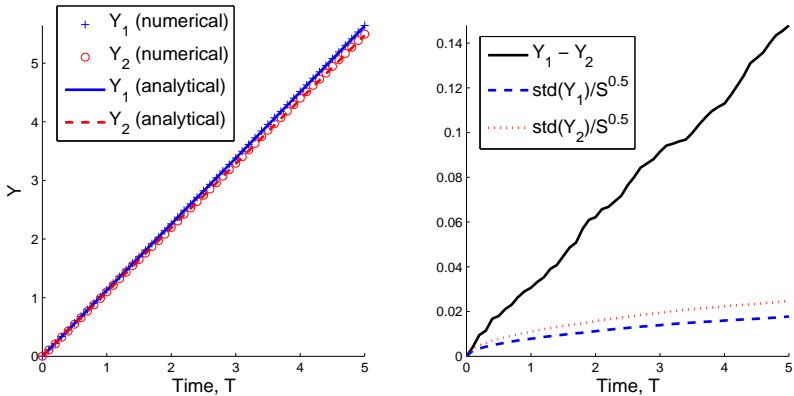}
\end{center}
\end{figure}

We now consider a two-risky-asset market model. With the model parameters given in \ref{simulations}, the optimal stock holdings in this case are
$A_1^* = 0.067 \times \Pi$ and $A_2^* = 0.467 \times \Pi$ (cf.\ \eqq (\ref{M2})). The performances of this optimal trading strategy based on our analytical expression in 
\eqq (\ref{analytical}) and our numerical simulations (cf.\ \ref{simulations} for details of simulation method) are given in \fig \ref{sec1}. If the correlation in the 
risky assets is ignored, the optimal portfolio becomes: $A_1^* = 0.3 \times \Pi$ and $A_2^* = 0.5 \times \Pi$. The corresponding performance is shown to be sub-optimal 
in \fig \ref{sec1}.

\section{Trading with transaction cost}
We will now include transaction cost into our discussion. 
As the transaction cost usually amounts to a small percentage ($\sim 0.5 \%$) of the total 
transaction, we employ perturbation method to analyse this optimisation problem with the transaction cost as the expansion parameter. By keeping track of the first few 
lowest order terms, we will derive the first order correction to the optimal trading strategy determined under no transaction cost. 

We assume that the transaction fee is proportional 
to the asset under transaction and the
proportionality constant is denoted by $k$. Note that we again define the total wealth, $\Pi$, as
\beq
\Pi = B +\sum_{i=1}^n A_i \ .
\eeq
The market model equations in this case are:
\begin{eqnarray}
\dd B & = & rB\dd t - (1+k) \dd L_i (t) + (1-k)\dd M_i (t) \nonumber
\\
&=& r\left[\Pi- (1-k)\sum^n_{i=1} A_i\right]\dd t - (1+k) \dd L_i (t) + (1-k)\dd M_i(t)
\nonumber
\\
\dd A_i & = & \mu _i A_i \dd t + \dd L_i (t) - \dd M_i (t) + \sigma_i A_i
\dd X_i\ ,\ \ i = 1,\cdots, n  \nonumber
\\
\dd \Pi & = & r\Pi+ (1-k)\sum^n_{i=1} \Big( -rA_i\dd t + \mu_i A_i \dd t + 
\sigma_i A_i \dd X_i \Big) \nonumber\\
&& -k \sum^n_{i=1} \Big( \dd L_i (t) + \dd M_i (t) \Big)
\end{eqnarray}
where $\q_i (t)$ and $\M_i (t)$ represent the cumulative purchase
and cumulative sale of assets $A_i$ during the time interval $[0,T]$.
The optimal expected value function $\cJ(\Pi, A, t)$ is as before:
\begin{equation}
\cJ(\Pi, A, t)  =  \max_{L_i, M_i } \e
\left[ \cF(\Pi(T)) + \int_{t}^{T} \cI(\Pi({t'})) d {t'} \right]\ ,
\end{equation}
and the corresponding HBJ equation is (Kamien 1991):
\begin{eqnarray}
0&=&
\max_{L_i, M_i} \Bigg\{ \cI +  \frac{\pp \cJ}{\pp t }+
\sum^n_{i=1} \left(\mu_i A_i + \frac{\dd L_i}{\dd t} - \frac{\dd M_i}{\dd t} \right) 
\frac{\pp \cJ}{\pp A_i}  
\nonumber
\\
&&
+\left[ r\left(\Pi - \sum^n_{i=1}A_i\right)+ \sum^n_{i=1} \left(\mu_i A_i - k \frac{\dd L_i}{\dd t} - k\frac{\dd M_i}{\dd t} \right)\right] \frac{\pp \cJ}{\pp \Pi} 
\nonumber
\\
&&
+  \sum^n_{i,j=1} \Omega_{ij}A_iA_j  \left(\frac{1}{2} \frac{\pp ^2 \cJ}{\pp A_i \pp A_j}+  \frac{1}{2}\frac{\pp ^2 \cJ}{ \pp \Pi^2}+ 
\frac{\pp ^2 \cJ}{\pp A_i \pp \Pi} \right) \Bigg\} \ .
\label{HBJ2}
\end{eqnarray}
Here, $L_i$ and $M_i$ are the control parameters from the dynamics programming perspective.

\subsection{Three regions}
By isolating terms involving $\dd L$ or $\dd M$ separately in \eqq (\ref{HBJ2}), we arrive at three separate cases:
\\
\\
{\it Case 1:} 
$\frac{\pp \cJ}{\pp A_i}-k\frac{\pp \cJ}{\pp \Pi}<0$ and $-\frac{\pp \cJ}{\pp A_i}-k\frac{\pp \cJ}{\pp \Pi} \geq 0$.
\\
In this case, the maximum in \eqq (\ref{HBJ2}) is achieved by choosing $\dd L_i=0$ and $\dd M_i=\infty$, which is equivalent to  selling at maximum rate.
\\
\\
{\it Case 2:}
$\frac{\pp \cJ}{\pp A_i}-k\frac{\pp \cJ}{\pp \Pi} \geq 0$ and $-\frac{\pp \cJ}{\pp A_i}-k\frac{\pp \cJ}{\pp \Pi} \leq 0$.
\\
In this case, the maximum is achieved by choosing $\dd L_i=\infty$ and $\dd M_i=0$, which is equivalent to buying at maximum rate.
\\
\\
{\it Case 3:}
$\frac{\pp \cJ}{\pp A_i}-k\frac{\pp \cJ}{\pp \Pi}<0$ and $-\frac{\pp \cJ}{\pp A_i}-k\frac{\pp \cJ}{\pp \Pi} < 0$.
\\
In this case, the maximum is achieved by choosing $\dd L_i=0$ and $\dd M_i=0$, which indicates that no transactions are needed.
\\
\\
\indent
We note that it is not possible to have $\frac{\pp \cJ}{\pp A_i}-k\frac{\pp \cJ}{\pp \Pi}$ and $-\frac{\pp \cJ}{\pp A_i}-k\frac{\pp \cJ}{\pp \Pi}$ be both greater than 
zero as we assume that $\cJ$ is an increasing function of $\Pi$.
This can be broadly interpreted as more wealth cannot decrease the value function from the trader's point of view.

With the above consideration, the optimal trading strategy can be seen to be partitioned into three separate 
regions: sale, purchase and no-transaction regions (cf.\ \fig \ref{pic}).
In other words, if the portfolio is in
the sale (purchase) region, the optimal strategy is to sell (buy) stocks until the
portfolio is at the no-transaction region boundary, and thus bring
the portfolio back into the no-transaction region. 
Inside the no-transaction region, $\dd L$ and $\dd M$ are identically zero and hence
$\cJ$ satisfies the HBJ equation with $k=0$.

\subsection{Continuity and optimality assumptions}
To make progress with our analysis, we will assume that the optimal value function, $J$, is everywhere continuous and that its derivatives are also continuous. We call 
the latter the optimality assumption.
The validities of these assumptions are discussed in Morton \& Pliska (1996) and Whalley \& Wilmott (1997).

 We now restrict ourselves to one risky asset for notational convenience. 
Suppose that the point $(\Pi, A, t)$ is inside the
sale region, when a very small quantity of assets, $h$ is sold, the
risk-free bond increases by the amount $h(1-k)$, while the whole
portfolio value is reduced by $kh$. As $h\rightarrow 0$, the value function $\cJ$ must be the same after the sale (the continuity assumption), we therefore have
\begin{eqnarray}
\lim_{h \rightarrow 0}\cJ(\Pi + kh, A, t) & = & \lim_{h \rightarrow 0}\cJ(\Pi , A-h, t)  
\\
\lim_{h \rightarrow 0}k \frac{\cJ(\Pi + kh,A, t)-\cJ(\Pi , A, t)}{kh}
&=&
\lim_{h \rightarrow 0}\frac{\cJ(\Pi , A-h, t)-\cJ(\Pi , A, t)}{h}
\\
\label{C1}
k \frac{\pp \cJ}{\pp \Pi}  &=&  -\frac{\pp \cJ}{\pp A} \ .
\end{eqnarray}
	
By a similar argument, we can conclude that inside the purchase region, we have
\beq
\label{C2}
k \frac{\pp \cJ}{\pp \Pi}  =  \frac{\pp \cJ}{\pp A} \ .
\eeq

By applying again the same argument to \eqs (\ref{C1}) and (\ref{C2}) with the use of the optimality assumption, we have that in the sale region and at the 
sale-no-transaction boundary:
\beq
\label{O1}
\frac{\pp ^2\cJ}{\pp A_i^2} = - k \frac{\pp \cJ}{\pp A_i \pp \Pi} \ ;
\eeq
and in the purchase region and at the purchase-no-transaction boundary:
\beq
\label{O2}
\frac{\pp ^2\cJ}{\pp A_i^2} =  k \frac{\pp \cJ}{\pp A_i \pp \Pi} \ .
\eeq

Inside the no-transaction region, the value function, $J$, must satisfy \eqq (\ref{HBJ2}) with $\dd L =\dd M =0$, i.e.,
\begin{eqnarray}
0&=&
 \cI +  \frac{\pp \cJ}{\pp t }+
\sum^n_{i=1} \mu_i A_i
\frac{\pp \cJ}{\pp A_i}  
+\left[ r\left(\Pi - \sum^n_{i=1}A_i\right)+ \sum^n_{i=1} \mu_i A_i \right] \frac{\pp \cJ}{\pp \Pi} \nonumber
\\
&&
+  \sum^n_{i,j=1} \Omega_{ij}A_iA_j  \left(\frac{1}{2} \frac{\pp ^2 \cJ}{\pp A_i \pp A_j}+  \frac{1}{2}\frac{\pp ^2 \cJ}{ \pp \Pi^2}+ 
\frac{\pp ^2 \cJ}{\pp A_i \pp \Pi} \right) \ .
\end{eqnarray}
These equalities are to be supplemented by the boundary condition at $t=T$:
$ \cJ(\Pi, A, T) = \cF(\Pi)$.

\subsection{Perturbative expansion and order matching}

We now redefine the $A_i$ coordinate as $A_i = A^*_i (\Pi, t) + k^{1/3} \alpha_i$,
where $A^*_i$ is the optimal value of stock $i$ held when $k$ tends to zero, and introduce the modified value function, $H$, such that $H(\Pi, \alpha, t) =
{\cJ}(\Pi, A, t)$. In \ref{appB}, we display the various derivatives of $J$ in terms of $H$ and $\alpha$. 

We further expand $H(\Pi, \alpha, t) $ in powers of $k^{1/3}$ as:
\begin{eqnarray}
&&
H_0(\Pi, \alpha, t) +k^{1/3}
H_1(\Pi, \alpha, t) +k^{2/3}H_2(\Pi, \alpha, t) 
\nonumber
\\
&+& kH_3(\Pi, \alpha, t) +k^{4/3}H_4(\Pi, \alpha, t) +{\cal O}(k^{5/3}) \ .
\label{expansion}
\end{eqnarray}
The reason for expanding $H$ and $A_i$ in powers of $k^{1/3}$ is out of necessity and has previously been studied in the literature (Atkinson \& Wilmott 1995, Rogers 
2004).

We will from now on keep track of the expression up to the first non-trivial correction: ${\cal O}(k^{5/3})$.
By matching the orders of $k$, equations (\ref{C1}) and (\ref{C2}) at the sale-no-transaction boundary (corresponds to the + sign in $\pm$) and at the 
purchase-no-transaction boundary (corresponds to the $-$ sign in $\pm$) become:
\begin{eqnarray}
0&=& \frac{ \pp H_m}{\pp \alpha_i}\sep 0 \leq m \leq 2
\label{C_1}
\\
0&=&\frac{ \pp H_3}{\pp \alpha_i} \pm \left(- \sum_{j=1}^n \frac{\pp A^*_j}{\pp \Pi} \frac{ \pp H_0}{\pp \alpha_j}\right)
\label{C_4}
\\
0&=&\frac{ \pp H_4}{\pp \alpha_i} \pm \frac{\pp H_0}{\pp \Pi} \pm \left( -\sum_{j=1}^n \frac{\pp A^*_j}{\pp \Pi} \frac{ \pp H_1}{\pp \alpha_j}\right) \ ,
\label{C_5}
\end{eqnarray}
and  equations (\ref{O1}) and (\ref{O2}) become:
\begin{eqnarray}
0&=&\frac{\pp^2 H_m}{\pp \alpha_i ^2} \sep 0 \leq m \leq 2
\label{S_1}
\\
0&=&\frac{\pp^2 H_3}{\pp \alpha_i^2} \pm \left(- \sum_{j=1}^n \frac{\pp A_j^*}{\pp \Pi} \frac{ \pp^2 H_0}{\pp \alpha_j^2}\right)
\label{S_4}
\\
0&=&\frac{\pp^2 H_4}{\pp \alpha_i^2}\pm \frac{\pp^2 H_0}{\pp \alpha_i \pp \Pi } \pm \left( -\sum_{j=1}^n \frac{\pp A_j^*}{\pp \Pi} \frac{ \pp^2 H_1}{\pp 
\alpha_j^2}\right) \ .
\label{S_5}
\end{eqnarray}
Inside the
no-transaction region, after 
expanding $H$ according to \eqq (\ref{expansion}) and collecting terms of
the same order in $k$, we arrive at the following conditions:

\begin{enumerate}
\item{${\cal O}(k^{-2/3})$ Equation:} $\cD H_0 = 0$,
where $\cD$ is an operator defined as $\sum^n_{i,j=1} D_{ij}\partial^2_{\alpha_i \alpha_j}$
with
\begin{equation}
\label{D11}
D_{ij} \equiv \frac{1}{2} \frac{\pp A^*_i}{\pp \Pi}\frac{\pp A^*_j}{\pp \Pi}\sum_{h,l=1}^n  \Omega_{hl}A^*_hA^*_l +  \frac{1}{2}\Omega_{ij} A^*_iA^*_j -
 \frac{\pp A^*_i}{\pp \Pi} \sum_{h=1}^n \Omega_{ih} A^*_iA^*_h
\end{equation}
\item{${\cal O}(k^{-1/3})$ Equation:}
$\cD H_1  =  0$.
\item{${\cal O}(1)$ Equation:} $\cD H_2  =  - \mm H_0$,
where $\mm$ is an operator defined as
\begin{equation}
\partial_t + \cI + r \left(\Pi -
\sum^n_{i=1} A^*_i \right) \partial_\Pi 
 +  \sum^n_{i=1}  \mu_i A^*_i  \partial_\Pi + \frac{1}{2}\sum_{i,j=1}^n
  \Omega_{ij} A^*_iA^*_j  \partial^2_{\Pi \Pi}  \ .
\end{equation}
\item{${\cal O}(k^{1/3})$ Equation:} 
$\cD H_3 =  - \sum^{n}_{i=1} \alpha_i \partial_{A_i}( \mm H_0 )- \mm H_1$.
\item{$O(k^{2/3})$ Equation:} $
\cD H_4  =  - \frac{1}{2}  \frac{\partial ^2 H_0}{\partial \Pi^2}\sum^{n}_{i,j=1}\Omega_{ij} \alpha_i \alpha_j
- \mm H_2$.
\end{enumerate}

Combining the ${\cal O}(k^{-2/3})$ equation with equations (\ref{C_1}) and 
(\ref{S_1}) when $m=0$, one finds
that $H_0$ is independent of $\alpha$.
Combining the ${\cal O}(k^{-1/3})$ with equations (\ref{C_1}) and (\ref{S_1}) when $m=1$ shows
that $H_1$ is independent of $\alpha$.
Combining the ${\cal O}(1)$ equation with equations (\ref{C_1}) and (\ref{S_1}) when $m=2$ shows 
that $H_2$ is independent of $\alpha$.
The ${\cal O}(k^{1/3})$ equation together with equations (\ref{C_4}) and (\ref{S_4}) imply 
that $H_3$ is independent of $\alpha$. 
In summary, by matching the coefficients of the various orders in $k$, we determine that $H_0,H_1,H_2,H_3$ are independent of $\alpha$.

Without loss of generality, we focus on the first asset and 
 let $\alpha_{1+}$ denotes the width of the purchase-no-transaction boundary, and
$\alpha_{1-}$ the width of the sale-no-transaction boundary (cf.\ \fig \ref{pic}).  From \eqs (\ref{C_5}) and (\ref{S_1}), we find that at the boundary 
$A^*+\alpha_{1-}$:
\beqn
\label{b+1}
\frac {\partial H_4}{\partial \alpha_{1}} + \frac {\partial
H_0}{\partial \Pi } &=& 0 
\\
\label{b+2}
 \frac {\partial^2 H_4}{\partial \alpha_{1}^2} &=&0 \ ,
\eeqn
and at boundary $A^*+\alpha_{1+}$, we have:
\beqn
\label{b-1}
\frac {\partial H_4}{\partial \alpha_1} - \frac {\partial
H_0}{\partial \Pi } &=& 0 
\\
\label{b-2}
 \frac {\partial^2 H_4}{\partial \alpha_1^2} &=&0
\ .
\eeqn
As we have established that $H_0$ and $H_2$ are independent of $\alpha$, with the ${\cal O}(k^{2/3})$ equation, we can conclude that $H_4$ has the following general 
form:
\begin{equation}
H_4(\Pi, \alpha, t)= \sum_{j=0}^4  h_j(\Pi, \alpha_{\bar{1}},t) \alpha_1^{j} \ ,
\eeq
where $\alpha_{\bar{1}}$ denotes the set $\{ \alpha_m : m>1\}$. In other words, $H_4$ is a polynomial in $\alpha_1$ with a degree of at most four.
 We now make the simplifying assumption that $\alpha_+ = -\alpha_-$. This is equivalent to saying that the transaction (buy or sell) boundaries are of the same distance 
away from the unperturbed optimal curve. We note that this assumption is proved to be true in the case of having uncorrelated risky assets (Atkinson and Mokkhavesa 
2004).
With the assumption of equal magnitude, we can conclude that $h_3=0$ at the boundaries by subtracting \eqq (\ref{b+2}) from \eqq (\ref{b-2}). In particular, we have
\beq
\label{h2}
6h_4 \alpha_+^2+h_2 =0 \ .
\eeq
By summing \eqs (\ref{b+1}) and (\ref{b-1}), we can further determine that $h_1=0$ at the boundaries.
By subtracting \eqq (\ref{b+1}) from \eqq (\ref{b-1}), we conclude that  $\alpha_{1+}$
satisfies:
\begin{equation}
\label{del_H0}
- \frac {\partial H_0}{\partial \Pi } =   4 h_4  \alpha_{1+}^3 +
2h_2 \alpha_{1+}  \ .
\end{equation}
Substituting \eqq (\ref{h2}) into \eqq (\ref{del_H0}), we obtain
\begin{equation}
\alpha_{1+}^3=  \frac{1}{8 h_4}  \frac {\partial H_0}{\partial \Pi }  \ . 
\label{eq:Hrelations}
\end{equation}
To calculate $h_4$, we invoke the ${\cal O}(k^{2/3})$ equation: By comparing the coefficient of the $\alpha_1^2$ term on both sides, we find that:
\begin{equation}
h_4=-\frac{\sigma^2_{1}}{24 D_{11}}
\frac{\partial ^2 H_0}{\partial \Pi^2} \ .
\end{equation}
So finally, $\alpha_{1\pm}$ can be expressed as:
\begin{equation}
\alpha_{1\pm}^3=  \mp \frac{3 D_{11}}{\sigma^2_1}  \frac {\partial H_0}{\partial \Pi }\left(\frac{\partial^2 H_0}{\partial \Pi^2}\right)^{-1} \ ,
\end{equation}
where $H_0$ is the optimal value function  when transaction cost is absent. 

In general, denoting the trading boundary for stock $i$ by $\alpha_{i+}$, we have the following general expression for the widths of the trading boundaries:
\begin{equation}
\label{width}
\alpha_{i+}=  \left| \frac{3 D_{ii}}{\sigma^2_i}  \frac {\partial H_0}{\partial \Pi }\left(\frac{\partial^2 H_0}{\partial \Pi^2}\right)^{-1}  \right|^{1/3} \ .
\end{equation}
For any financial model where $H_0$ is known, the above equation together with \eqq (\ref{A}) provides an analytical description of the optimal trading strategy.
This is the main result of this paper.

\subsection{Example: the Long Term Growth Model}
According to \eqs \ref{M1} and \ref{M2}:
\beqn
H_0(\Pi,t)&=& \log \Pi + \left(r +\frac{1}{2} \hat{\mu} \cdot \Omega^{-1} \hat{\mu} \right) (T-t)
\\
A^*&=&\Pi \Omega^{-1} \hat{\mu} \ .
\eeqn
Combining these with \eqq (\ref{D11}), we have
\beqn
D_{ii} &=& \frac{1}{2} \frac{\pp A^*_i}{\pp \Pi}\frac{\pp A^*_j}{\pp \Pi}\sum_{h,l=1}^n  \Omega_{hl}A^*_hA^*_l +  \frac{1}{2}\Omega_{ij} A^*_iA^*_j -
 \frac{\pp A^*_i}{\pp \Pi} \sum_{h=1}^n \Omega_{ih} A^*_iA^*_h
\\
&=& \Pi^2 \left\{ \frac{1}{2} (\hat{\mu} \cdot \Omega^{-1} \hat{\mu} +\sigma_i^2) [\Omega^{-1} \hat{\mu} ]_i^2
- \hat{\mu}_i  [ \Omega^{-1} \hat{\mu} ]_i^2
\right\} \ .
\eeqn
The width of the boundary for stock $i$ is therefore (cf. \eqq (\ref{width})):
\begin{equation}
\label{M3}
\Pi
\left\{
\frac{3k}{\sigma_i^2}
\left[\frac{1}{2} (\hat{\mu} \cdot \Omega^{-1} \hat{\mu} +\sigma_i^2) [\Omega^{-1} \hat{\mu} ]_i^2
- \hat{\mu}_i  [ \Omega^{-1} \hat{\mu} ]_i^2\right] \right\}^{1/3}  \ .
\end{equation}
If the risky assets are uncorrelated, the expression above coincides with the result of Atkinson \& Mokkkhavesa (2004).

\begin{figure}
\caption{
Simulation results for the two-risky-asset market with transaction cost. {\it Left plot:} $Y_1 \equiv \la \log(\Pi(T)) \ra$ denotes the performance of the optimal 
trading strategy, while $Y_2$ denotes the performance of the sub-optimal strategy obtained if the correlation between the two risky assets is ignored in the calculations 
for the boundary widths.
{\it Right plot:} The plots of $(Y_1 - Y_2)$ and the standard errors of the means versus time based on our simulations. Note that $S = 15000$ is the number of samples in 
the simulations.
}
\label{sec2a}
\begin{center}
\includegraphics[scale=.45]{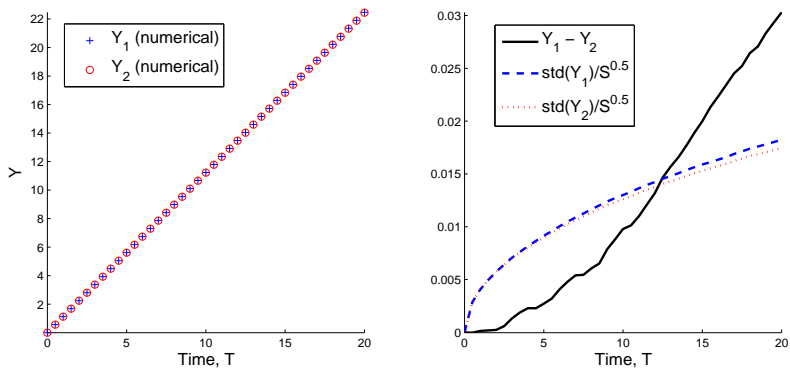}
\end{center}
\end{figure}
\begin{figure}
\caption{
A particular simulation run with the optimal trading strategy in the two-risky asset model.  {\it Upper plot:} the temporal evolution of values held in bond and stocks. 
{\it Lower plot:} The transactions performed according to the optimal trading strategy in the time interval $t \in [4, 6]$. The purchases (sales) of stock $A_1$ are 
denoted by blue (green) crosses, and the purchases (sales) of stock $A_2$ are denoted by black (red) triangles. Note that the transaction amount is not infinitesimal 
only because of the discrete time evolution (tick time) in the simulations (cf.\  \ref{simulations}).
}
\label{sec2b}
\begin{center}
\includegraphics[scale=.45]{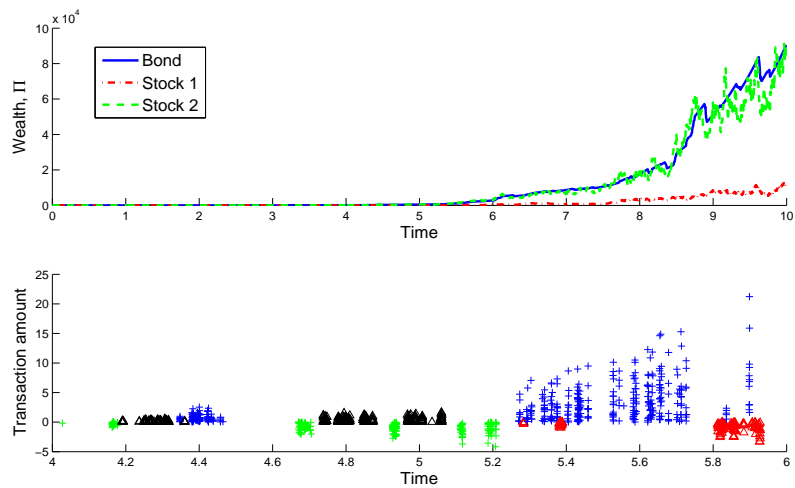}
\end{center}
\end{figure}

We now employ this optimal trading strategy to the two-risky-asset market considered before. The optimal curve corresponds to: $A_1^* = 0.067 \times \Pi$ and $A_2^* = 
0.467 \times \Pi$ (cf. \eqq (\ref{M2})), and according to \eqq (\ref{M3}), the boundaries widths are: $\alpha_{1+} = 0.167 \times k^{1/3} \Pi$ and $\alpha_{2+} = 0.710 
\times k^{1/3} \Pi$. The performance of this strategy is shown in \fig \ref{sec2a}. If we ignore the correlation between the risky assets in 
calculating the boundary widths, $\alpha_{1+}$ and $\alpha_{2+}$ become $0.508 \times k^{1/3} \Pi$ and $0.760 \times k^{1/3} \Pi$ respectively. The trading strategy 
employing these boundaries together with the same optimal curve as before is shown in \fig \ref{sec2a} and can be seen to be sub-optimal, albeit the difference is small. 

In \fig \ref{sec2b}, the portfolio's temporal evolution of a particular simulation is shown together with the transaction amounts displayed.

\section{Conclusion}
In conclusion, we have employed perturbation method to study multi-asset optimisation for arbitrary utility functions. By making the assumption that the sale and 
purchase boundaries are of the same distance away from the optimal curve, we arrived at an analytical expression for the optimal trading strategy. We have also supported 
our analytical results with numerical simulations in the context of the long term growth model.

\appendix{Details of numerical simulation method}
\label{simulations}

We consider a portfolio consisting of two risky assets and one risk-free asset. 
The values in the bond and risky assets are updated as follows:
\beqn
 B (t+\tri t) &=& B(t) + r B(t) \tri t
 \\
 A_1(t+\tri t) &=& A_1(t) + \mu_1 A_1(t) \tri t + \sigma_1 A_1(t) \sqrt{\tri t} \ z_1(t)
 \\
 A_2(t+\tri t) &=& A_2(t) + \mu_2 A_2(t) \tri t + \sigma_2 A_2(t)  \sqrt{\tri t} \ z_2(t)
\eeqn
 where $z_1(t)$ and $z_2(t)$ are random numbers drawn from the normal distribution with zero mean and a standard deviation of one, such that the correlation coefficient 
between $z_1(t)$ and $z_2(t)$ is $\rho_{12}$.

In the case of trading without transaction costs, the portfolio is updated after each iteration according to \eqq (\ref{M2}). When transaction costs are present, trading 
only occurs when the value of the risky assets are outside of the no-transaction region (cf.\ \fig \ref{pic}), i.e., if
\beq
A_i(t) \notin [A_i^*(t) +\alpha_{i+}(t) , A_i^*(t)+\alpha_{i-}(t)] \ .
\eeq
 When such an event occur, the portfolio is adjusted such that $A_i(t)$ is moved back to the nearest boundary and the cost of transaction is subtracted from the wealth. 
For example, if $A_1(t)> A_i^*(t) +\alpha_+(t)$, then the portfolio is adjusted so that: 
 \beqn
 B(t) &\rightarrow & B(t) +(1-k) [A_1(t)-A_1^*(t) -\alpha_{1+}(t)]
 \\
  A_1(t) &\rightarrow & A_1^*(t) +\alpha_{1+}(t) \ .
 \eeqn
 
The simulations always start with a total wealth of 1 at the optimal portfolio distribution and the set of parameters employed are: $r=1, \mu_1=1.3, \mu_2=1.5, \sigma_1 
=\sigma_2=1, \rho_{12} =0.5, k=0.005$ and $\tri t = 5\times 10^{-5}$.

\appendix{Change of variables}
\label{appB}
Letting $H(\Pi, \alpha, t) = J(\Pi, A,t)$ with $A = A^*(\Pi,t) +k^{1/3} \alpha$, we have the following expressions for the derivative of $J$ in terms of $H$ and 
$\alpha$.
\beqn
\frac{\pp J}{\pp A_i} &=& k^{-1/3} \frac{\pp H}{ \pp \alpha_i}
\\
\frac{\pp J}{\pp \Pi} &=& \frac{\pp H}{\pp \Pi} -\sum_{i=1}^n k^{-1/3} \frac{\pp H}{ \pp \alpha_i}\frac{\pp A^*_i}{ \pp \Pi}
\\
\frac{\pp J}{\pp t} &=& \frac{\pp H}{\pp t} -\sum_{i=1}^n k^{-1/3} \frac{\pp H}{ \pp \alpha_i}\frac{\pp A^*_i}{ \pp t}
\\
\frac{\pp^2 J}{\pp A_i \pp A_j} &=& k^{-2/3} \frac{\pp^2 H}{ \pp \alpha_i \pp \alpha_j}
\\
\nonumber
\frac{\pp^2 J}{\pp \Pi^2} &=& \frac{\pp^2 H}{\pp \Pi^2} -k^{-1/3}\sum_{i=1}^n  \left(2
 \frac{\pp^2 H}{ \pp \alpha_i \pp \Pi}\frac{\pp A^*_i}{ \pp \Pi} +
 \frac{\pp H}{ \pp \alpha_i }\frac{\pp^2 A^*_i}{ \pp \Pi^2}  \right)
 \\
 && +k^{-2/3} \sum_{i,j=1}^n 
 \frac{\pp^2 H}{ \pp \alpha_i \pp \alpha_j}\frac{\pp A^*_i}{ \pp \Pi} \frac{\pp A^*_j}{ \pp \Pi} 
 \\
 \frac{\pp^2 J}{\pp \Pi \pp A_i} &=& k^{-1/3}\frac{\pp^2 H}{\pp \Pi \pp \alpha_i} -k^{-2/3}\sum_{j=1}^n \frac{\pp^2 H}{ \pp \alpha_i \pp \alpha_j}\frac{\pp A^*_j}{ \pp 
\Pi} 
 \ .
\eeqn

\subsection*{Acknowledgements}
We would like to thank Colin Atkinson and Michael Giles for valuable comments.
SLL thanks the Croucher Foundation, CFL thanks the Glasstone Trust (Oxford) and Jesus College (Oxford), for financial support.

\end{document}